\documentclass{desyproc_rrnumber}

\begin{document}
\title{Dark Matter Decay and Cosmic Rays\footnote{Talk presented at the
\textit{5th Patras Workshop, Durham, UK, 13-17 July 2009.}}}

\author{{\slshape Christoph Weniger}\\[1ex]
DESY, Notketra{\ss}e 85, 22607 Hamburg, Germany}

\contribID{lindner\_axel}

\desyproc{DESY-PROC-2009-05}
\acronym{Patras 2009} 

\maketitle

\begin{abstract}
  The decay of dark matter is predicted by many theoretical models and can
  produce observable contributions to the cosmic-ray fluxes. I shortly discuss
  the interpretation of the positron and electron excess as observed by PAMELA
  and Fermi LAT in terms of decaying dark matter, and I point out the
  implications for the Fermi LAT observations of the $\gamma$-ray flux with
  emphasis on its dipole-like anisotropy.
\end{abstract}

\section{Introduction}
The most popular type of dark matter (DM) candidate, the weakly interacting
massive particle (WIMP), can naturally reproduce the observed DM abundance due
to effective self-annihilation in the early Universe, and today this same
annihilation process could produce an observable contribution to the measured
cosmic-ray fluxes on Earth. Such an indirect detection of DM is also possible
if DM \textit{decays} with a sufficiently large rate. There exist a number of
interesting DM models (see \textit{e.g.}~\cite{models} and references therein)
that predict the decay of DM on cosmological time scales, namely with
lifetimes around and above $\tau_\text{DM}\simeq
\mathcal{O}(10^{26}\,\text{s})$, which are typically required to be not in
conflict with current observational limits. Among these models is the
gravitino with a small violation of $R$-parity, motivated by requiring a
consistent thermal history of the Universe, and the sterile neutrinos, whose
long lifetime is due to tiny Yukawa couplings. The typical masses for these DM
candidates lie in the $100\,\text{GeV}$ and the $10\,\text{keV}$ regime,
respectively. Another interesting model with kinetically mixed hidden gauginos
was also recently studied~\cite{hg}. Even in models where DM is stable in the
first place, the consideration of higher-dimensional operators often renders
the DM particle unstable with cosmological lifetimes. Since the indirect
detection signals from decay differ in general from the ones of annihilation,
a dedicated study of decaying DM signals is mandatory. Below I will shortly
review the $\gamma$-ray and $e^\pm$-signals that can come from DM decay, and I
will discuss them in light of recent observations.

\section{Cosmic rays from dark matter decay}
Provided the decays occur at a sufficiently large rate, their products could
be observable as an exotic contribution to the high energy cosmic ray fluxes
of $\gamma$-rays, electrons, positrons, antiprotons, neutrinos or
antideuterons.  Among the different cosmic-ray species, $\gamma$-rays play a
distinct role, due to their sensitivity to far-distant sources and their
potential to discriminate between astrophysical and DM signals. The
\textit{gamma-ray} signal from DM decay consists of several components. The
most important one is related to the prompt radiation (\textit{e.g.}~final
state radiation) produced in the decay of DM particles inside the Milky Way
halo. It depends on the halo density profile, and although the halo profile is
expected to be approximately isotropic, the corresponding flux at Earth
exhibits a strong dipole-like anisotropy due to the offset between sun and
galactic center. In contrast, the extragalactic prompt component of the
$\gamma$-ray signal, which stems from the decay of DM particles at
cosmological distances, is largely isotropic. At energies around
$10\,\text{GeV}$ or below, the magnitude of the  halo and extragalactic fluxes
are of the same order, whereas at much higher energies around $1\,\text{TeV}$
the inelastic scattering between $\gamma$-rays and the intergalactic
background light renders the extragalactic component negligible.  Decaying DM
in general also produces electrons and positrons, which give rise to another
contribution of the $\gamma$-ray signal, coming from the inverse Compton
scattering (ICS) between the electrons and positrons and the interstellar
radiation field (ISRF). This component is highly anisotropic and usually lower
in energy than the component from prompt radiation.  The main background in
the $\gamma$-ray channel is the diffuse emission of our Galaxy, which is
mainly due to interactions of cosmic rays with the galactic gas and the ISRF.
This component is by far strongest in the galactic disk region, and it turns
out that exotic fluxes from DM decay would dominantly show up at higher
latitudes, away from the disk. This is in contrast to annihilation signals,
which are often expected to be best seen very near to the galactic center.

The \textit{electrons and positrons} produced in the Milky Way halo by DM
decay scatter on irregularities of the Galactic magnetic field, which results
in a wash-out of directional information before they reach the Earth. Their
propagation is commonly described by a diffusion model, whose free parameters
are tuned to reproduce the observed cosmic-ray nuclei fluxes. The
astrophysical background in this channel in mainly due to primary electrons,
which are presumably produced in supernova remnants, and due to secondary
positrons, produced in the interaction of cosmic-rays with the galactic gas.

Recently it has become apparent that state-of-the-art propagation models fail
to reproduce the PAMELA measurements of the \textit{positron fraction} at
energies larger than $10\,\text{GeV}$~\cite{Adriani:2008zr}. Together with the
more recent Fermi LAT and HESS data \cite{FermiHESS} for the total
$e^\pm$-flux the experiments suggest an excess of electrons and positrons up
to energies around a few $1\,\text{TeV}$. The most common astrophysical
explanation of this excesses is the electron-positron pair production by the
interactions of high-energy photons in the strong magnetic field of nearby
pulsars, such as Geminga or Monogem (see \textit{e.g.}~\cite{pulsar} and
references therein). However, an arguably more exciting explanation of the
cosmic-ray electron/positron excesses is the possibility that the electrons
and positrons are produced in the annihilation or the decay of DM particles.

\section{Positron excess and gamma-ray prospects}
\begin{figure}[ht]
  \vspace{-.2cm}
  \begin{center}
    \includegraphics[width=\linewidth]{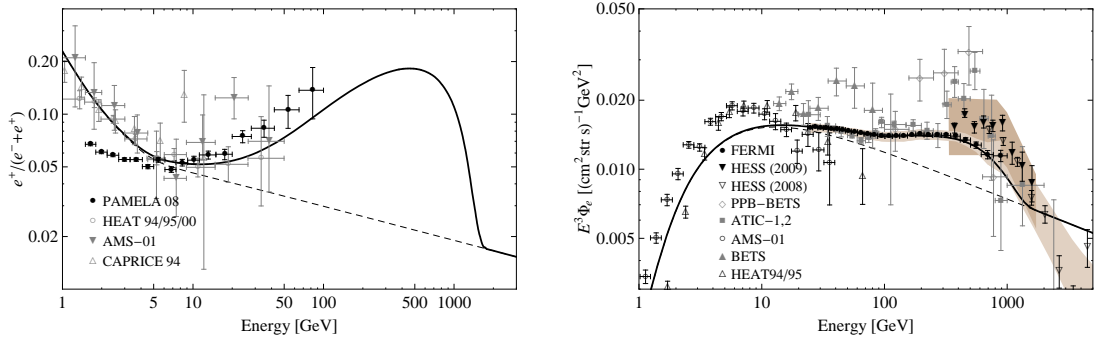}
  \end{center}
  \vspace{-.5cm}
  \caption{Positron fraction (\textit{left panel}) and electron+positron flux
  (\textit{right panel}) for DM decay $\psi_\text{dm}\rightarrow
  \mu^+\mu^-\nu$ (see Tab.~\ref{tab:results}).  The dashed line shows the
  astrophysical background. Details are given in Ref.~\cite{Ibarra:2009dr}.}
  \label{fig:mumunu}
\end{figure}

\begin{wraptable}{r}{8cm}
  \centering
  \begin{tabular}{|c|c|c|}
    \hline
    Decay Channel & $M_\text{DM}$ [GeV] & $\tau_\text{DM}$
    [$10^{26}$s]\\\hline
    $\psi_\text{DM}\rightarrow \mu^+\mu^-\nu$ & 3500 & 1.1 \\
    $\psi_\text{DM}\rightarrow \ell^+\ell^-\nu$ & 2500 & 1.5\\
    $\phi_\text{DM}\rightarrow \mu^+\mu^-$ & 2500 & 1.8 \\
    $\phi_\text{DM}\rightarrow \tau^+\tau^-$ & 5000 & 0.9 \\\hline
    $\psi_\text{DM}\rightarrow W^\pm\mu^\mp$ & 3000 & 2.1 \\\hline
  \end{tabular}
  \caption{DM decay channels that we found to best fit the Fermi LAT and
  PAMELA data~\cite{Ibarra:2009dr}.}
  \label{tab:results}
\end{wraptable}
If the observed excess of positrons and electrons is entirely due to DM decay,
one obtains clear predictions for the $\gamma$-ray signal that should be
observable at Fermi LAT.  In Ref.~\cite{Ibarra:2009dr} we analyzed the
predictions for the positron fraction and the total electron plus positron
flux including a possible contribution from DM decay in order to account for
the anomalies observed by PAMELA and Fermi.  We considered several scenarios
of decaying DM, being it either a fermionic or a bosonic particle, which
decays into various channels with a branching ratio of 100\%. Our results are
summarized in Tab.~\ref{tab:results}, an example is shown in
Fig.~\ref{fig:mumunu}. From the data leptonic, and in particular muonic, modes
are favored. Note that the decay into $W^\pm\mu^\mp$ is in some tension with
the anti-proton/proton ratio observed by PAMELA.\\

\begin{figure}[ht]
  \vspace{-.2cm}
  \begin{center}
    \includegraphics[width=\linewidth]{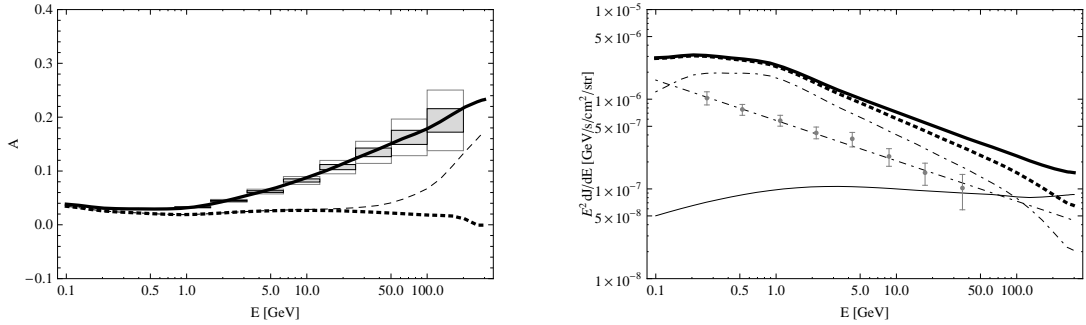}
  \end{center}
  \vspace{-.5cm}
  \caption{\textit{Left panel:} Anisotropy of $\gamma$-ray signal for the
  decay mode $\psi_\text{dm}\rightarrow\mu^+\mu^-\nu$~\cite{Ibarra:2009nw}.
  The \textit{solid} line shows the anisotropy of the total signal, including
  the galactic foreground, the \textit{dotted} line shows the anisotropy of
  the foreground alone. The \textit{thin dashed} line is the anisotropy of the
  total signal when neglecting ICS radiation of electrons and positrons from
  DM decay. \textit{Right panel:} Averaged fluxes of the different
  $\gamma$-ray components. Line coding as in left panel, in addition the
  \textit{thin solid} line shows the pure DM signal and the
  \textit{dot-dashed} lines show the adopted extragalactic background flux and
  the galactic foreground. Data points correspond to the preliminary Fermi LAT
  results for the extragalactic $\gamma$-ray background~\cite{Talk}.}
  \label{fig:ani}
\end{figure}

The production of electrons and positrons in the DM decay inevitably produces
also contributions to the cosmic $\gamma$-rays. In particular the
$\gamma$-ray signal of the decay modes shown in Tab.~\ref{tab:results} should
give rise to a clear signal in the Fermi LAT observations at higher latitudes.
Furthermore this signal is expected to be anisotropic, which can be used to
discriminate it from the galactic foreground and the extragalactic
$\gamma$-ray background. To illustrate this we define the anisotropy parameter
$A=( {\bar{J}_\text{GC}-\bar{J}_\text{GAC}})/(
{\bar{J}_\text{GC}+\bar{J}_\text{GAC}})$, where $\bar{J}_\text{GC}$ and
$\bar{J}_\text{GAC}$ denote the diffuse $\gamma$-ray flux averaged over the
hemisphere in direction of the Galactic center (GC) and anticenter (GAC),
respectively. The galactic disk, with latitudes $|b|<10^\circ$, is excluded
form the average. The left panel of Fig.~\ref{fig:ani} shows our results
for the anisotropy parameter which is expected to be observed by the Fermi LAT
if the DM particle decays into $\mu^+\mu^-\nu$ (see Ref.~\cite{Ibarra:2009nw}
for details). Although the decay channel is marginally consistent with
preliminary data (right panel), a sizeable anisotropy, around
$A\simeq0.2-0.3$, is predicted at energies $E_\gamma\simeq100\,\text{GeV}$.
This can be significantly different from the anisotropy of the astrophysical
foreground (we adopt the conventional model \texttt{44\_500180} from
\texttt{galprop.stanford.edu}). As indicated by our estimates of the
statistical error bars for one-year and five-year Fermi LAT observation, this
deviation should be clearly visible in the upcoming results for the diffuse
$\gamma$-ray sky. 

\section{Conclusions}
Many theoretical models predict the decay of DM on cosmological timescales,
giving rise to an anomalous contribution to the observed cosmic-ray fluxes.
The corresponding $\gamma$-ray signals could show up as broad features over
large angular distance in the $\gamma$-ray sky. If decaying DM is the right
explanation of the positron and electron excess observed by PAMELA and Fermi
LAT, a corresponding $\gamma$-ray signal with a large dipole-like anisotropy
should be observed in the very near future with Fermi LAT. This anisotropy
would be due to prompt radiation at high latitudes, and due to ICS radiation
at lower latitudes, most prominent in a region of a few kpc around the
galactic center. It is tempting to speculate that such an ICS signal already
showed up in the Fermi LAT data, see Ref.~\cite{Dobler:2009xz}.

\section*{Acknowledgments}
The author likes to thank the organizers of the \textit{5th Patras Workshop on
Axions, WIMPs and WISPs} for an enlightening conference, and Alejandro Ibarra
and David Tran for very fruitful collaboration.

\begin{footnotesize}

\end{footnotesize}
\end{document}